\documentclass[preprint2]{aastex}
\usepackage{aastexug}
\newcommand{\pks}{PKS~1055+201}

\newcommand{\axaf}{\mbox{\em Chandra\/}}

\usepackage{graphicx}
\usepackage{color}

\begin{document}

\title{Discovery of an X-ray Jet and Extended Jet Structure in the
  Quasar PKS~1055+201}
\author{D.A. Schwartz,\altaffilmark{1} 
H.L. Marshall,\altaffilmark{2} J.E.J. Lovell,\altaffilmark{3} D.W.
Murphy,\altaffilmark{4} G.V. Bicknell,\altaffilmark{5} 
M.~Birkinshaw,\altaffilmark{1,6}
J.M.~Gelbord,\altaffilmark{2}  M. Georganopoulos,\altaffilmark{7,8}
L. Godfrey,\altaffilmark{3,5}
D.L. Jauncey,\altaffilmark{3}  
S. Jester,\altaffilmark{9}
E.S.~Perlman,\altaffilmark{7}
D. M. Worrall\altaffilmark{1,6} }
\altaffiltext{1}{Harvard-Smithsonian Center for Astrophysics} 
\altaffiltext{2}{Kavli Institute for Astrophysics and Space Research,
  MIT} 
\altaffiltext{3}{CSIRO Australia Telescope National Facility} 
\altaffiltext{4}{Jet Propulsion Laboratory} 
\altaffiltext{5}{Research School of Astronomy and Astrophysics,
  ANU}
\altaffiltext{6}{Department of Physics, University of Bristol} 
\altaffiltext{7}{Department of Physics, 
  University of Maryland-Baltimore County}
\altaffiltext{8}{NASA's Goddard Space Flight Center}
\altaffiltext{9}{School of Physics and Astronomy, University of
  Southampton}

\email{das@head-cfa.harvard.edu}

\begin{abstract}

This letter reports rich X-ray jet structures found in the \axaf\
observation of PKS 1055+201. In addition to an X-ray jet coincident
with the radio jet we detect a region of extended X-ray emission
surrounding the jet as far from the core as the radio hotspot to the
North, and a similar extended X-ray region along the presumed path of
the unseen counterjet to the Southern radio lobe.  Both X-ray regions
show a similar curvature to the west, relative to the quasar. We
interpret this as the first example where we separately detect the
X-ray emission from a narrow jet and extended, residual jet plasma
over the entire length of a powerful FRII jet.

\end{abstract}

\keywords{(galaxies:) quasars: general, quasars: individual (PKS
  1055+201=4C 20.24), galaxies: jets, X-rays: galaxies}

\section{INTRODUCTION}

The surprising discovery of a luminous, 100-kpc-scale X-ray jet in the
initial observations of PKS~0637--752 \citep{Schwartz00, Chartas00}
demonstrated that \axaf\ had opened a new channel of data for the
study of extragalactic jets. One of the extreme cases discovered in
our survey for X-ray emission from radio-bright quasar jets
\citep{Marshall05a, Gelbord05a} is the FR II radio source
\object[PKSB1055+201]{PKS 1055+201} \citep{Shimmins68}. Those authors
associated it with the previously discovered 4C20.24
\citep{Pilkington65}. \citet{Bolton68} identified the object as a
quasar at redshift z=1.11. We proposed \axaf\ followup observations
based on our initial 4.7 ks image
(\dataset[ADS/Sa.CXO#obs/04889]{obsid 4889}) which showed that the
X-ray jet followed the 21\arcsec\ length of the radio jet.  At the
redshift of the quasar, 1\arcsec =8.24 kpc in the plane of the
sky\footnote{We use a flat, accelerating cosmology, with H$_0$=71 km
s$^{-1}$ Mpc$^{-1}$, $\Omega_{m} = 0.27$ , and
$\Omega_{\Lambda}=0.73$.}. Since we infer that the jet is in
relativistic motion with Doppler factor $\delta \approx $6
\citep{Schwartz06b}, the jet must be at an angle no more than 9\degr\
from our line of sight, so that its total deprojected length is of
order 1 Mpc.

In this letter we discuss the extended X-ray emission surrounding the
prominent radio and X-ray jet and the inferred, invisible
counter-jet. The ability to observe both the jet and X-ray
tube\footnote{We use \emph{tube} rather than \emph{lobe} or
\emph{cocoon} to avoid pre-judging the nature of this region.} may
help constrain such problems as how the energy of a jet is dissipated
and the jet decelerated
\citep[e.g.,][]{Georganopoulos03, Georganopoulos04, Tavecchio06}, or
how the radio lobes relate to the X-ray jet. Details of the structure
of the narrow jet \citep{Schwartz06b}, and of our \emph{HST}
observations of the system, will be presented elsewhere.

\section{IMAGES OF \pks\ }

\begin{figure}[t]
\includegraphics[width=.96\columnwidth]{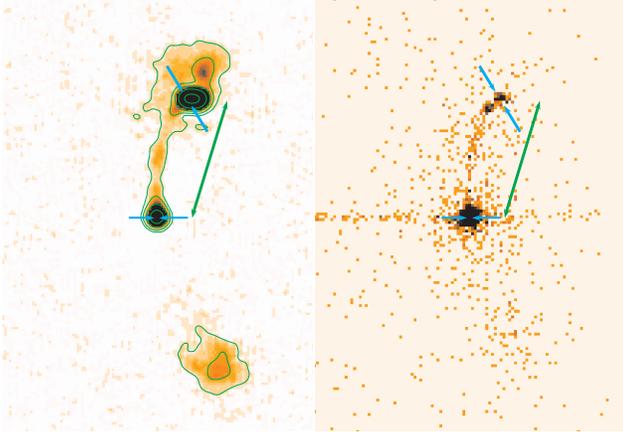}
 \caption{\label{fig:images} X-ray and radio images of \pks . Right
 panel plots the 0.5 to 7 keV X-ray data, in 0.49\arcsec\ bins. Left
 panel shows a 1.46 GHz VLA A-array image, restored with a 1\farcs65
 $\times$ 1\farcs50 beam. Both images are to the same scale, with the
 21\arcsec --long arrow indicating the length of the X-ray jet. Radio
 contours are factors of 4 starting from 1.5 mJy/beam, with a peak of
 768 mJy/beam. The rms noise is 0.6 mJy/beam. The quasar and terminal
 X-ray hotspot are marked with pairs of arrows. The horizontal streak
 through the quasar in the X-ray image is the ACIS readout artifact. }
\end{figure}

\axaf\ \dataset[ADS/Sa.CXO#obs/05733]{obsid 5733} yielded a 31.595 ks
live-time observation of \pks.  Figure~\ref{fig:images} presents radio
and X-ray images of the system \pks.  The radio jet to the north
closely resembles the X-ray jet up to the terminal X-ray hotspot about
21\arcsec\ from the quasar. That hotspot is associated with the
eastern peak of a bright, resolved radio hotspot. A radio extension to
the north of this hotspot is not reproduced in the X-ray image.  Both
the radio and X-ray jet start out at a position angle within a few
degrees of North, close to the VLBA jet direction of -5\degr\ reported
by \citet{Kellerman04}. They reported superluminal motion with
an apparent velocity $v=(10\pm4)c$, for a component 3 mas from the
core.

\begin{figure} [t]
\includegraphics[viewport=20 10 410 300,clip,
width=.98\columnwidth]{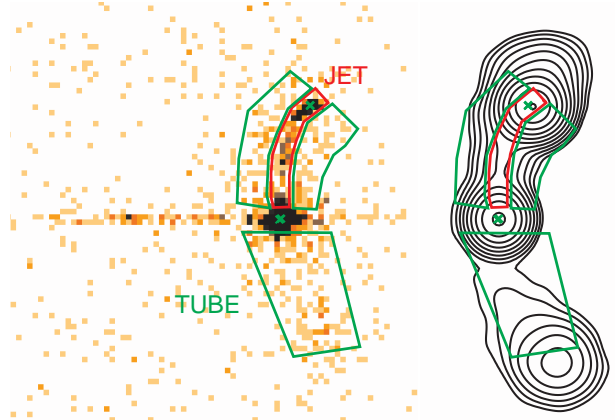} 
\caption{\small \label{fig:regions} Left panel: X-rays plotted in
1\arcsec bins. The narrow X-ray jet is denoted by the region enclosed
in red. The extended X-ray ``tube'' of emission is indicated by the 3
regions enclosed in green. Right panel shows 1.425 GHz VLA-B contours,
which are factors of 2 times 1.2mJy/beam, with 1.54 Jy/beam peak. The
rms noise is 0.4 mJy/beam. The south tube region flux is about 12 mJy,
and the north region is consistent with a similar flux plus the narrow
jet smeared by the 5'' beam FWHM. Scales are the same in both panels,
as are the ``tube'' and ``jet'' regions. The same points indicated in
figure 1 are here marked with an ``X''. }
\end{figure}

Binning the X-ray data (0.5 to 7 keV) into 1\arcsec\ bins shows
enhanced X-ray emission from a region about 16\arcsec\ wide and
parallel to the radio and X-ray jet to the North
(Figure~\ref{fig:regions}).  A roughly symmetric region of enhanced
X-rays also appears to the south.  Its position suggests that it
surrounds the unseen counterjet, which powers the southern radio lobe.

\begin{figure}[t]
\plottwo{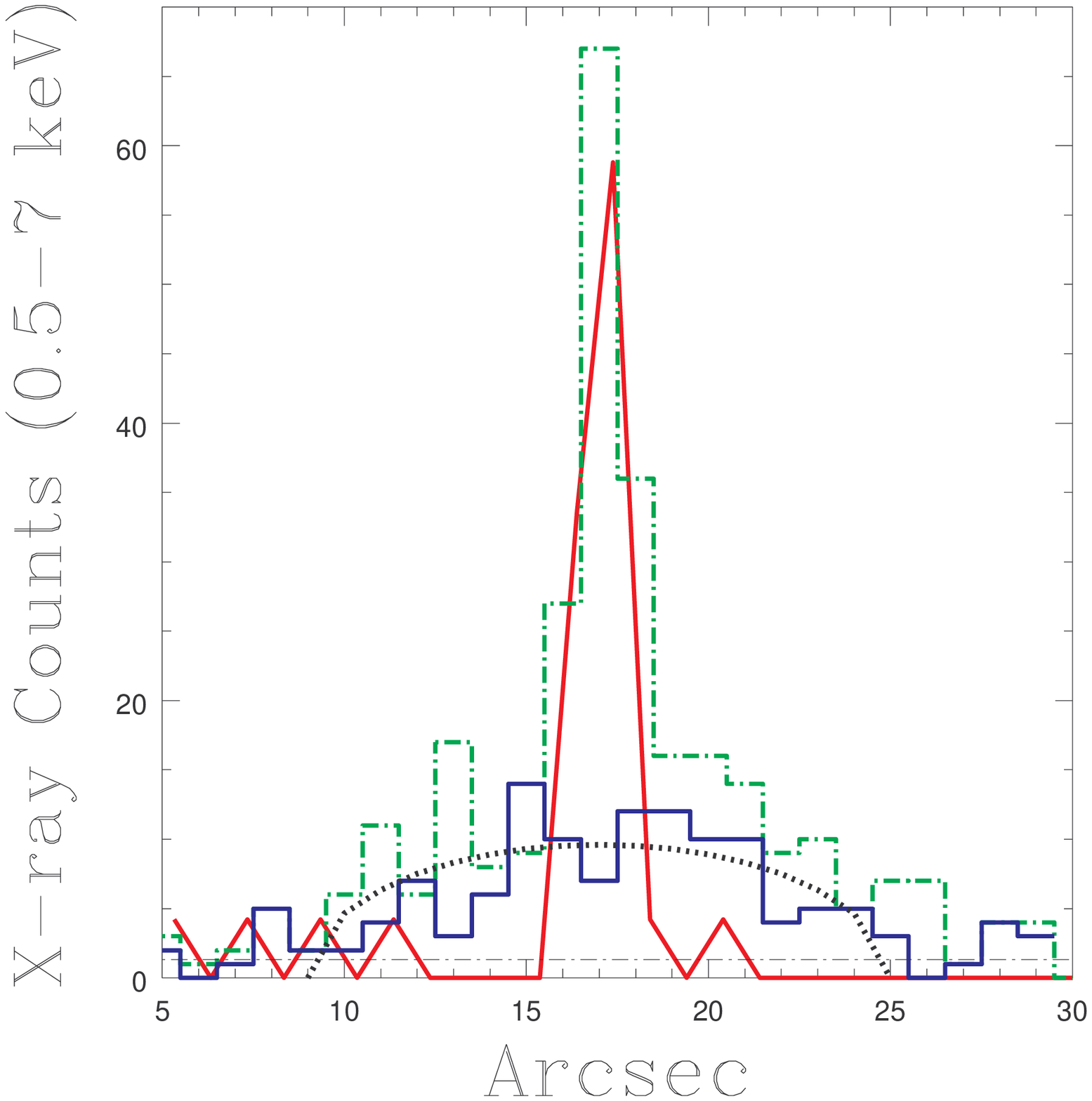}{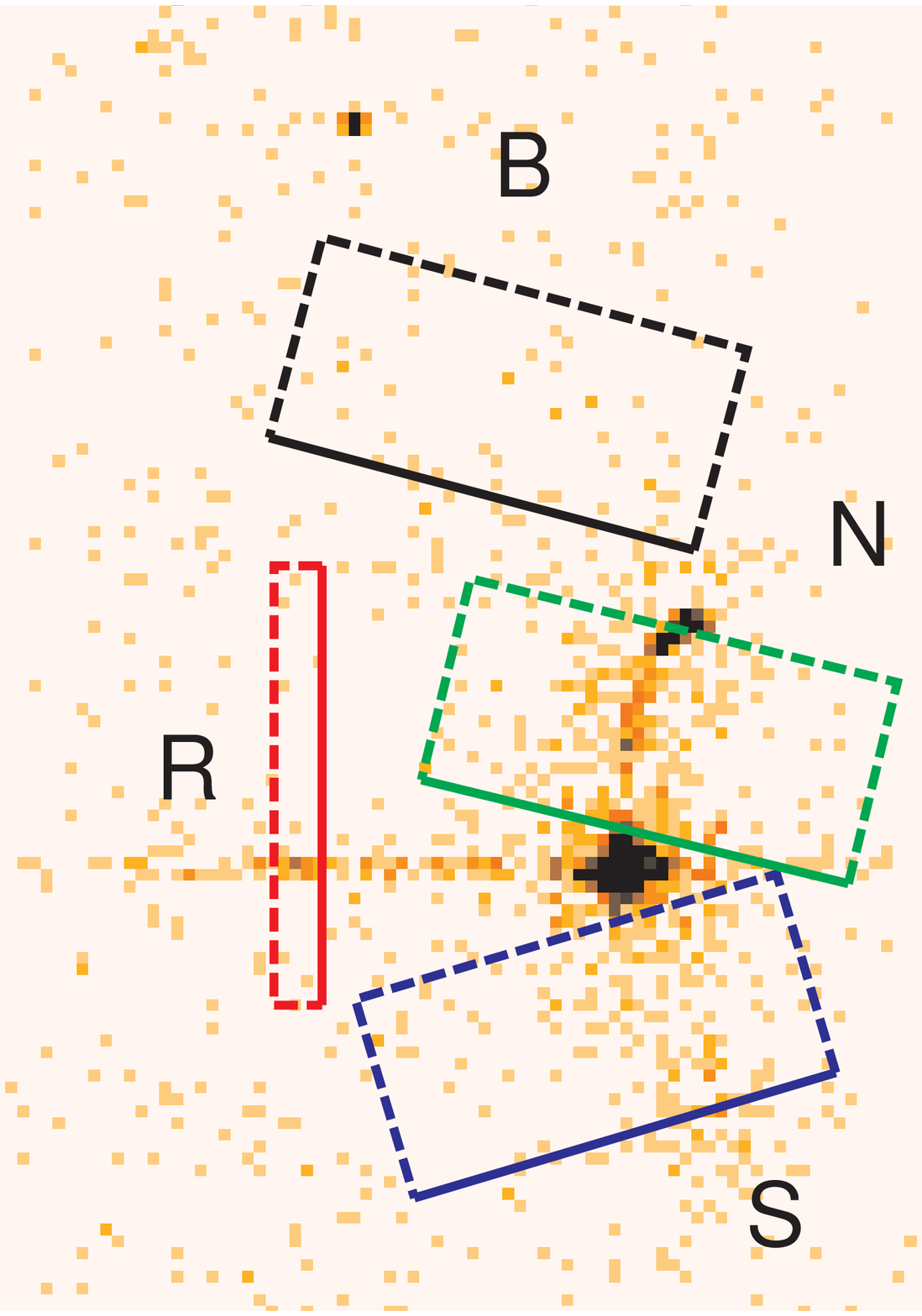}
 \caption{\label{fig:profiles}The left panel shows the profiles of
 projections across the regions defined in the X-ray image in the
 right panel. Green dot-dashed histogram shows the northern jet and tube
 (``N''), blue solid histogram shows the southern tube (``S'' ), the
 dashed line gives the background as determined from region ``B'', and
 the red line gives the profile across the readout streak, (``R'').
 The heavy dotted line models the extended X-ray profiles as from a
 uniform cylinder.  The projection collapses counts within the dashed
 rectangles into 1\arcsec\ bins along each solid line. We plot from
 left to right along that line of projection, offsetting the profiles
 so that the centroids are aligned. }
\end{figure}

We quantify the extended X-ray emission by considering the profiles in
Figure~\ref{fig:profiles}. The top panel gives the profiles across the
corresponding regions shown in the bottom panel. The X-ray counts are
the sum of the data projected into bins 1 arcsec wide along the solid
lines in the bottom panel. The abscissae are arbitrarily offset to
line up the centroids of the emission in each case. The dashed line
gives the background rate expected, 1.31 counts per bin, based on the
region marked \emph{B}. The red line gives the profile across the ACIS
readout streak from the bright quasar core. This represents the undersampled
profile of an unresolved line source. The actual counts in the streak
profile are multiplied by 4.2 to give the number of counts which the
narrow jet is expected to contribute to the profile (green histogram)
from the North region\footnote{Since the jet curves a small amount,
its contribution to the profile is slightly broader than the line
response function given by the readout streak.}. The South region
profile is given by the blue histogram. Errors on the tube region
profiles are the square root of the number of counts in the bin. The
dotted curve is the profile predicted for a uniformly emitting
cylinder of diameter 16\arcsec.

We can make some simple estimates by modeling the extended X-ray
brightness as from a uniformly emitting cylinder. We consider the
observed energy range 0.5 to 7 keV, and subtract 0.0685 counts arcsec$^{-2}$
background to derive all the numbers quoted. From the measured counts in the
regions outside the jet (green areas in Figure~\ref{fig:regions}) we
estimate that 47.5 of the 243.8 background-subtracted counts in the jet
region (red area in in Figure~\ref{fig:regions}) are due to projection
of the extended tube region in front and behind the jet. We then infer
that there are 196.3 $\pm$ 14.1 counts intrinsic to the jet, and 175.6
$\pm$ 14.0 and 145.2 $\pm$ 13.7 counts in the North and South tube
regions, respectively.  Using the spectral
models described in the next section, we estimate rest frame 2--10 keV
luminosities of approximately 1.8 $\times$ 10$^{44}$ ergs s$^{-1}$ for
the jet, and 1.8 $\times$ 10$^{44}$ ergs s$^{-1}$ and 1.5 $\times$
10$^{44}$ ergs s$^{-1}$ for the North and South tube regions,
respectively.

\section{X-RAY SPECTRA}

We use XSPEC \citep{Arnaud96} to fit the X-ray spectra of the
extended X-ray tube emission and of the jet, where those two regions are
defined in Figure~\ref{fig:regions}. In all cases we fix the Galactic
absorption to column density N$_{\rm H}=1.9\times 10^{20} $cm$^{-2}$
\citep{Stark92}. We take the energy regions 0.5--5 keV, (since there
are few counts between 5 and 7 keV), bin the data
into 25 counts or more, and use a $\chi^2$ goodness of fit estimator. We first
fitted the spectrum of the background to a power law. Then we fitted the
spectrum of the extended X-ray emission to the sum of the normalized,
fixed background fit plus either an absorbed power law, or absorbed
thermal (MEKAL) model. Then we fitted the spectrum of the jet to the sum
of the normalized, fixed background fit plus the normalized, fixed
fit to the extended region for each of the two cases, plus an absorbed
power law. We quote 90\% confidence error regions for one interesting
parameter, either a temperature or power-law  photon index, $\Gamma$.

The background is fitted by a power-law $\Gamma$=0.54 $\pm$
0.23. Background contributes an estimated 12\% of the extended-region
spectrum, and 2\% of the jet-region spectrum. The MEKAL fit to the
extended region gives $kT= 8.6^{+34}_{-4.5} $ keV with a $\chi^2$ of
14.05 for 8 degrees of freedom, where we have fixed the abundances at
cosmic values. The power-law fit for this region gives $\Gamma = 1.73
\pm 0.33$ with a $\chi^2$ of 9.4. The power law is clearly a better
fit to the extended region, with a probability $P(>\chi^2)$=0.31, but
the thermal fit cannot be rejected since $P(>\chi^2)$=0.08, and it is
interesting to interpret both models. The jet power-law index is
$\Gamma = 2.15 \pm$0.33 with $\chi^2$ of 2.33 for 6 degrees of freedom
if the MEKAL fit is assumed for the extended region, and $\Gamma =
2.11 \pm$ 0.33 with $\chi^2$ of 2.38 if the power-law fit is assumed.
The spectral shape deduced for the jet is not sensitive to the model
of the extended region, and we shall consider only the parameters
obtained from assuming the power-law fit to the extended
emission. Such a fit for the jet region has an $P(>\chi^2)$=0.88.

The $\Gamma$ values for the jet and tube regions are consistent at the
90\% confidence level.  A joint absorbed power-law fit of the extended
and jet regions, forcing the photon index to be the same in both,
results in an acceptable $\chi^2$ of 13.81 for 15 degrees of freedom,
with $\Gamma$=1.92$\pm$0.19, to 90\% confidence.

\section{DISCUSSION}
\label{sec:discussion}

 We model the jet assuming minimum energy conditions between the
magnetic field and relativistic particles, with equal electron and
proton energy densities. We assume the X-ray emission is inverse
Compton (IC) scattering from the cosmic microwave background (CMB),
which requires that the jet be relativistically beamed near to our
line of sight \citep{Tavecchio00,Celotti01}. Details of these
assumptions are the same as in \citet{Schwartz06a}. We derive mean jet
parameters \citep{Schwartz06b} which are roughly as follows: Magnetic
field in the rest frame $B \approx$ 10$\mu$G (1 nT = 10$\mu$G);
electron density in the rest frame $n_e \approx 2 \times 10^{-8} {\rm
cm^{-3}}$ (between Lorentz factors $\gamma$ from 80 to 10$^5$ and with
an assumed radio energy index of $\alpha$=0.7); Doppler factor $\delta
\approx$ 6; kinetic flux $\approx$ 2 $\times 10^{45}$ ergs s$^{-1}$;
angle to the line of sight $\theta \le$ 9\degr; and deprojected length
$\approx$ 1 Mpc when we assume the maximum $\theta$=9\degr. These
quantities become $B \approx 60 \mu$G, $n_e \approx $10$^{-6} {\rm
cm^{-3}}$, and $\delta \approx $4.5 at the terminal X-ray hotspot.

We consider two mechanisms for the extended X-ray emission: thermal
bremsstrahlung from gas heated by the jet or IC/CMB from relativistic
electrons whose ultimate energy source is from the jet. We consider
the total volume, V, of two cylinders, each 16\arcsec\ = 130 kpc in
diameter by 1 Mpc in length, and emitting a total 3.3 $\times 10^{44}$
ergs s$^{-1}$ in the rest frame 2--10 keV band. We assume that the
cylinder is not moving relativistically with respect to the CMB, since
the two sides of the source are of similar length, and that the X-ray
emissivity is apparently uniform throughout the cylinder volumes.

If the extended X-ray emission is thermal and uniform, then $kT = 8.6$
keV and $n_e = 0.004$ cm$^{-3}$.  The cooling time would be $1.2
\times 10^{10}$ yr, much longer than the age of the universe at $z =
1.1$, $t = 5.5 \times 10^{9}$ yr.  The total energy would be $E = 3
n_e kT V \approx 1.6 \times 10^ {62}$ erg.  Assuming that there is a
counterjet of similar power, the jets supply a kinetic flux of $4
\times 10^{45}$ erg/s, requiring $1.3 \times 10^{9}$ yr to supply the
total energy.  A duty cycle of 25\% seems high but could be reduced if
the gas is non-uniform.  Then the $<n_e>$ and $E$ could be $\sim 10
\times$ smaller while maintaining $<n_e^2> = (0.0043)^2$ cm$^{-6}$ in
order to produce the observed X- ray luminosity. Alternatively, one
might infer that previous outbursts of the jet averaged a kinetic flux
which was a factor of several more luminous than the jet we presently
observe, or that the present jet has a ratio of proton to electron
energy a factor of several higher than the value unity which we
assume.  The confinement of such a gas is problematical in the absence
of a rich cluster medium, for whose presence we have no
indication. Nonetheless, we could be observing a system similar to Cyg
A \citep{Wilson06}, but tilted closer to our line of sight and
substantially larger. Measurement of the rotation measure might test
for the presence of a thermal tube, or of a hot cluster medium.  If
magnetic fields of the order of 0.1 $\mu$G were present in such an
extended medium, we would expect a rotation measure RM $\approx$ 20
$\times$ (B$_{\|}$/0.1$\mu$G).

In the case of non-thermal emission of X-rays from the extended
region, the low level of 1.4 GHz radio emission implies that the
magnetic field is much lower than 14 $\mu$G, which is the field which
would have the same energy density as the CMB at the redshift z=1.11. 
With an upper limit of 10mJy for the radio emission, the equipartition
field would be about 6 $\mu$G. We will assume the X-ray
emission is IC/CMB from a region \emph{not} in bulk relativistic
motion with respect to the microwave background. Under those
conditions, we deduce a density of electrons with Lorentz factor above
$\gamma$=80 of 2.2 $\times 10^{-8} {\rm
cm^{-3}}$, comparable to the density within
most of the length of the jet.  The electron lifetimes are limited by
scattering on the CMB, and are of order 10$^8$ years for electrons producing
inverse Compton radiation at 1 keV, but less than 10$^7$ years for
the higher energy electrons emitting 1.46 GHz synchrotron radiation.

The non-thermal scenario is consistent with an interpretation wherein
the jet is generating X-ray emitting lobes throughout a lifetime of
several times 10$^7$ years, and these lobes are decelerated to
substantially less than the speed of light, while the jet continues to
propagate with a large Lorentz factor. To diffuse laterally to the
observed radial extent of 65 kpc in the 10$^8$ year lifetime of the
X-ray emitting electrons implies a transverse velocity of order 600 km
s$^{-1}$. While radio lobes are ubiquitous, and while X-ray emission
has been detected from radio lobes \citep[e.g.,][and references
therein]{Croston05}, the present observation of \pks\ is unique in
that both a prominent X-ray jet and a surrounding X-ray tube region
are seen, and that the latter does not correlate morphologically with
the weak, extended radio emission at 1.46 GHz.

\acknowledgments This work was supported by NASA contract NAS8-39073
to the \emph{Chandra} X-ray Center, and SAO SV1-61010 to MIT, and NASA
grant GO2-3151C to SAO. JMG was supported by NASA grants GO4-5124X and
GO5-6116A.  ESP acknowledges support from NASA LTSA grant
NAG5-9997. SJ was supported by the MPI f\"ur Astronomie through an
Otto Hahn fellowship. Part of this research was performed at the Jet
Propulsion Laboratory, California Institute of Technology, under
contract to NASA.  We thank M. Hardcastle, and D. Harris for
discussions and comments. This research used the NASA Astrophysics
Data System Bibliographic Services, and the NASA/IPAC Extragalactic
Database (NED) which is operated by the Jet Propulsion Laboratory,
California Institute of Technology, under contract with the NASA.


\begin{thebibliography}{}

\bibitem[Arnaud(1996)]{Arnaud96} Arnaud, K. A. 1996, in
  A.S.P. Conference Series, Vol. 101, ``Astronomical Data Analysis
  Software and Systems V'', eds. Jacoby, George H. \& Barnes,
  Jeannette 17

\bibitem[Bolton, Kinman, \& Wall(1968)]{Bolton68} 
Bolton, J. G., Kinman, T. D., \&  Wall, J. V. 1968, \apj, 154, L105
\bibitem[Celotti et al.(2001)]{Celotti01} Celotti, A., Ghisellini,
G., \& Chiaberge, M. 2001, \mnras, 321, L1, astro-ph/0008021
\bibitem[Chartas et al.(2000)]{Chartas00} Chartas, G., et al. 2000,
\apj, 542, 655, astro-ph/0005227
\bibitem[Croston et al.(2005)]{Croston05} Croston, J. H., Hardcastle,
  M. J., Harris, D. E., Belsole, E., Birkinshaw, M., \& Worrall,
  D. M. 2005, \apj, 626, 733, astro-ph/0503203
\bibitem[Gelbord et al.(2005)]{Gelbord05a}Gelbord, J., et al. 2005,
  in ``X-Ray and Radio Connections '', eds. L.O. Sjouwerman and K.K
  Dyer, (NRAO: http://www.aoc.nrao.edu/events/xraydio),  astro-ph/0411804
\bibitem[Georganopoulos \& Kazanas(2003)]{Georganopoulos03}
  Georganopoulos, M. \& Kazanas, D. 2003, \apj, 589, L5
\bibitem[Georganopoulos \& Kazanas(2004)]{Georganopoulos04}
  Georganopoulos, M. \& Kazanas, D. 2004, \apj, 604, L81
\bibitem[Kellerman et al.(2004)]{Kellerman04} Kellerman, K. I., et
  al. 2004, \apj, 609, 539
\bibitem[Marshall et al.(2005)]{Marshall05a} Marshall, H. L., et al.
2005, \apjs, 156, 13
\bibitem[Pilkington \& Scott(1965)]{Pilkington65}Pilkington, J.~D.~H.,
  \& Scott, J.~F.\ 1965, \memras, 69, 183 
\bibitem[Schwartz et al.(2000)]{Schwartz00} Schwartz,
D. A., et al. 2000,\apjl, 540, L69
\bibitem[Schwartz et al.(2006a)]{Schwartz06a} Schwartz, D. A., et
al. 2006a, \apj, 640, 592
\bibitem[Schwartz et al.(2006b)]{Schwartz06b} Schwartz, D. A. et
al. 2006b, in Proc. "The X-ray Universe 2005", San Lorenzo
de El Escorial, Spain, September 2005, 579 
\bibitem[Shimmins \& Day(1968)]{Shimmins68} Shimmins, A.~J., \&
Day, G.~A.\ 1968, Australian Journal of Physics, 21, 377
\bibitem[Stark et al.(1992)]{Stark92} Stark, A. A., et al 1992, ApJS 79. 77
\bibitem[Tavecchio et al.(2000)]{Tavecchio00} Tavecchio, F.,Maraschi,
L., Sambruna, R. M., Urry, C. M.  2000, \apjl, 544, L23
\bibitem[Tavecchio et al.(2006)]{Tavecchio06} Tavecchio, F., Maraschi,
  L., Sambruna, R. M., Gliozzi, M., Cheung, C. C., Wardle, J. F. C.,
  Urry, \& C. Megan  2006, \apj, 641, 732
\bibitem[Wilson, Smith \& Young(2006)]{Wilson06} Wilson, A. S., Smith,
  David A., \& Young, A. J. 2006, \apjl, 644, L9
\end{thebibliography}
\end{document}